\documentclass[aps,prl,twocolumn,groupedadress,nofootinbib]{revtex4-1}
\usepackage{graphicx}
\usepackage{dcolumn}
\usepackage{bm}
\usepackage{dsfont,amsmath,amssymb,natbib,color,units,pstricks}

\newcommand \be  {\begin{equation}}
\newcommand \bea {\begin{eqnarray} \nonumber }
\newcommand \ee  {\end{equation}}
\newcommand \eea {\end{eqnarray}}

\begin{document}

\title{Eigenvector dynamics: theory and some applications \\}

\author{Romain Allez$^{1,2}$ and Jean-Philippe Bouchaud$^1$}
\affiliation{$^1$ Capital~Fund~Management, 6--8 boulevard Haussmann, 75\,009 Paris, France}
\affiliation{$^2$ Universit{\'e} Paris-Dauphine, Ceremade, 75\,016 Paris, France.}
\date{\today}
\begin{abstract}
We propose a general framework to study the stability of the subspace spanned by $P$ consecutive eigenvectors of a generic symmetric matrix ${\bf H}_0$, when a small perturbation is added. This problem is relevant in various contexts, including quantum dissipation (${\bf H}_0$ is then the Hamiltonian) 
and risk control (in which case ${\bf H}_0$ is the assets return correlation matrix). We specialize our results for the case of a Gaussian Orthogonal 
${\bf H}_0$, or when ${\bf H}_0$ is a correlation matrix. We illustrate the usefulness of our framework using financial data.  
\end{abstract}

\maketitle

Random Matrix Theory (RMT) is extraordinary powerful at describing the eigenvalues statistics of large random, or pseudo-random, matrices \cite{Verdu,Handbook}. Eigenvalue 
densities, two-point correlation functions, level spacing distributions, etc. can be characterized with exquisite details. The ``dynamics'' of 
these eigenvalues, i.e. the way these eigenvalues evolve when the initial matrix ${\bf H}_0$ is perturbed by some small matrix $\epsilon {\bf P}$, is also well understood \cite{Simon}. 
The knowledge 
of the corresponding eigenvectors is comparatively much poorer. One reason is that many RMT results concern rotationally invariant matrix ensembles, 
such that by definition the statistics of eigenvectors is featureless. Still, as we will show below, some interesting results can be derived for the
dynamics of these eigenvectors. Let us give two examples for which this question is highly relevant.

One problem where the evolution of eigenvectors is important is quantum dissipation \cite{Wilkinson}. As the parameters of the Hamiltonian ${\bf H}_t = {\bf H}_0 + 
\epsilon {\bf P}_t$ of a system evolve with time $t$, the average energy changes as well. One term corresponds to the average (reversible) change 
of the Hamiltonian that leads to a shift of the energy levels (the eigenvalues). But if the external perturbation is not infinitely slow, some 
transitions between energy levels will take place, leading to a dissipative (irreversible) term in the evolution equation of the average energy 
of the system. The adiabaticity condition which ensures that no transition takes place amounts to comparing the speed of change of the perturbation 
$\epsilon {\bf P}_t$ with a quantity proportional to the typical spacing between energy levels. For systems involving a very large number $N$ of 
degrees of freedom, the average level spacing of the $N \times N$ Hamiltonian ${\bf H}$ goes to zero as $N^{-1}$. For $N \to \infty$, any finite speed of change therefore
corresponds to the ``fast'' limit, where a large number of transitions between states is expected. In fact, if the quantum system is in state 
$|\phi_i^0 \rangle$ at time $t=0$, that corresponds to the $i$th eigenvector of ${\bf H}_0$, the probability to jump to the $j$th eigenvector of ${\bf H}_1$,  
$|\phi_j^1 \rangle$, at time $t=1$ is given by $|\langle \phi_j^1 | \phi_i^0 \rangle|^2$, where we use the bra-ket notation for vectors and scalar
products. The way energy is absorbed by the system will therefore be determined by the perturbation-induced distortion of the eigenvectors. More precisely, 
if $|\phi_i^0 \rangle$ is different from $|\phi_i^1 \rangle$, some transitions must take place in the non-adiabatic limit, that involve
all the states $j$ which have a significant overlap with the initial state.

Another very relevant situation is quantitative finance, where the covariance matrix ${\bf C}$ between the returns of $N$ assets (for example stocks) plays a
major role in risk control and portfolio construction \cite{Review}. More precisely, the risk of a portfolio that invests $w_\alpha$ in asset $\alpha$ is given by 
${\mathcal R}^2 = \sum_{\alpha\beta} w_\alpha {\bf C}_{\alpha\beta} w_\beta$. Constructing low risk portfolios requires the knowledge of the $n$ largest eigenvalues of ${\bf C}$, 
$\lambda_1 \leq \dots \leq \lambda_n$ and their corresponding eigenvectors $|\phi_1 \rangle, \dots, |\phi_n \rangle$. A portfolio such that the vector of weights $| w  \rangle$ 
has zero overlap with the first $n$ eigenvectors of ${\bf C}$ has a risk that is bounded from above by $\lambda_{n+1}$. The problem with this idea is that
it relies on the assumption that the covariance matrix ${\bf C}$ is perfectly known and constant in time. The observation of a sufficiently long time
series of past returns would allow one, in such a stable world, to determine ${\bf C}$ and to immunize the portfolio against risky investment modes. 
Unfortunately, this idea is thwarted by two (inter-related) predicaments: a) time series are always of finite length, and lead to substantial 
``noise'' in empirical estimates of ${\bf C}$ \cite{Review} and b) the world is clearly not stationary and there is no guarantee that the covariance matrix 
corresponding to the pre-crisis period 2000-2007 is the same as the one corresponding to the period 2008-2011. For one thing, some companies disappear 
and others are created in the course of time. But even restricted to companies that exist throughout the whole period, it is by no means granted that
the correlation between stock returns do not evolve in time. This is why it is common practice in the financial industry to restrict the period
used to determine the covariance matrix to windows of a few years into the recent past. This leads to the measurement noise problem alluded above. 
Now, if the ``future'' large eigenvectors do not coincide with the past ones, a supposedly low risk portfolio will in fact be exposed to large 
risks directions in the future. Denoting as $|\phi_i^0 \rangle$ the past eigenvectors and $|\phi_j^1 \rangle$ the future ones, the ``unexpected risk'' of
portfolio $|\phi_i^0 \rangle $ can be defined as $\sum_{j=1}^n \lambda_j^1 \langle \phi_j^1 |\phi_i^0 \rangle^2$. Therefore, 
as for the quantum dissipation problem, the statistics of the overlaps $G_{ij} = \langle \phi_i^1 | \phi_j^0 \rangle$ is a crucial piece of information.  

When trying to follow the evolution of a given eigenvector $|\phi_i \rangle$ as the perturbation is increased, one immediately faces a problem when
eigenvalues ``collide''. It is well known that true collisions (degeneracies) are non generic; the collisions are in fact avoided and levels do not
cross. However, upon the pseudo-collision of $\lambda_i$ and $\lambda_{i+1}$ (say), the eigenvectors $|\phi_i \rangle$ and $|\phi_{i+1} \rangle$ 
strongly hybridize. Therefore, single eigenvectors are strongly unstable in time. The idea is then to study the stability of the subspace spanned by 
$2p+1$ several consecutive eigenvalues: $\left\{|\phi_{k-p}^0\rangle, \dots |\phi_{k}^0\rangle, \dots, |\phi_{k+p}^0\rangle\right\}$. Motivated by the 
above examples, we ask the following question: how should one choose $q \geq p$ such that the subspace spanned by the set
$\left\{|\phi_{k-q}^1\rangle, \dots |\phi_{k}^1\rangle, \dots, |\phi_{k+q}^1\rangle\right\}$ has a significant overlap with the initial subspace? In order
to answer this question, we consider the $(2q+1) \times (2p+1)$ rectangular matrix of overlaps ${\bf G}$ with entries $G_{ij}$. The $(2p+1)$ non zero singular 
values $1 \geq s_1 \geq s_2 \dots s_{2p+1} \geq 0$ of ${\bf G}$ give full information about the overlap between the two spaces. For example, 
the largest singular value $s_1$ indicates that there is a certain linear combination of the $(2q+1)$ perturbed eigenvectors that has a scalar product
$s_1$ with a certain linear combination of the $(2p+1)$ unperturbed eigenvectors. If $s_{2p+1}=1$, then the initial subspace is entirely spanned by the
perturbed subspace. If on the contrary $s_1 \ll 1$, it means that the initial and perturbed eigenspace are nearly orthogonal to one another. A good 
measure of the distance $D$ between the two spaces is provided by $-\langle \ln s \rangle=-(\sum_i \ln s_i)/(2p+1)$ (but alternative measures, such as $1-\langle s \rangle$, could be considered as well). 
Since the singular values $s$ are obtained as the square-root of the eigenvalues of the matrix ${\bf G}^\dagger{\bf G}$,
one has $D(p,q) \equiv - \ln \det {\bf G}^\dagger{\bf G}/2P$, where we introduce for convenience the notations $P=2p+1$, $Q=2q+1$. When two subspaces of dimensions $P$ and $Q$ 
are constructed using randomly chosen vectors in a space of dimension $N$, one expects accidental overlaps, such that $D(p,q)$ is in fact finite. 
The distance can be calculated exactly using Random Matrix Theory tools in the limit $N,P,Q \to \infty$, with $\alpha=P/N$ and $\beta=Q/N$ held fixed. The result is \cite{RSVD}:
\begin{equation*}
D_{RMT} = - \int_0^1 {\rm d} s \ln(s) \frac{\sqrt{(s^2-\gamma_-)_+ (\gamma_+-s^2)_+}}{\beta\pi s (1-s^2)}
\end{equation*}
where $\gamma_\pm = \alpha+\beta-2\alpha\beta \pm 2\sqrt{\alpha\beta(1-\alpha)(1-\beta)}$. In other words, in that limit, the full density of singular 
values is known. This provides a benchmark to test whether the two eigenspaces are accidentally close ($D \approx D_{RMT}$), or if they are genuinely 
similar ($D \ll D_{RMT}$).

Endowed with the above formalism, we can now proceed to compute $D$ in the case where the perturbation is small. Using standard perturbation theory, the
perturbed eigenvectors can be expressed in terms of the initial eigenvectors as:
\begin{align*}
|\phi^1_i\rangle &= {\left(1- \epsilon^2\sum_{j \neq i} \left(\frac{P_{ij}}{\lambda_i - \lambda_j} \right)^2 \right)}^{1/2} |\phi_i^0 \rangle 
+ \epsilon \sum_{j \neq i} \frac{P_{ij}}{\lambda_i - \lambda_j} |\phi_j^0\rangle \\ 
&+  \epsilon^2 \sum_{j \neq i} \frac{1}{\lambda_i - \lambda_j} \left( \sum_{\ell \neq i} \frac{P_{j\ell} P_{\ell i}}{\lambda_i - \lambda_\ell} - 
\frac{P_{ii}P_{ij}}{\lambda_i - \lambda_j} \right) | \phi_j^0 \rangle,
\end{align*}
where $P_{ij} \equiv  \langle \phi_j^0|{\bf P}|\phi_i^0 \rangle$.
The denominators $\lambda_i - \lambda_j$ remind us that eigenvectors are strongly affected by eigenvalue pseudo-collisions, as alluded to above. The above expression allows one to obtain the overlap matrix ${\bf G}$. 
Keeping only the relevant terms to order $\epsilon^2$, one finds:
\begin{equation}
G_{ij} = \begin{cases} 
1 - \frac{\epsilon^2}{2} \sum_{\ell \neq i} \left(\frac{P_{i\ell}}{\lambda_i - \lambda_\ell} \right)^2 & \text{if $i=j$},\\
\epsilon \frac{P_{ij}}{\lambda_i - \lambda_j}  &  \text{if $i \neq j$}. 
\end{cases} 
\end{equation}
It is then easy to derive the central result of our study: to second order in $\epsilon$, the distance $D(p,q)$ between the initial and perturbed eigenspaces is:
\be
D(p,q) \approx  \frac{\epsilon^2}{2P} \sum_{i=k-p}^{k+p} \sum_{j\notin \{k-q, \dots, k+q\}} \left(\frac{P_{ij}}{\lambda_j - \lambda_i} \right)^2.
\ee
We now turn to two explicit illustrations, first in the context of the GOE matrices, and then in the context of empirical correlation matrices.

{\it{Eigenvector stability in the GOE ensemble.}} We now choose $H_0$ and $P$ to be two independent realizations of the Gaussian Orthogonal Ensemble 
of random matrices of size $N \times N$. We normalize the elements of $H_0$ and $P$ to have variance $1/N$, such that the density of eigenvalues $\rho(\lambda)$ tends to the Wigner semi-circle when $N \to \infty$.
We consider the subspace of initial eigenvectors corresponding to all the eigenvalues $\lambda$ contained in a certain finite interval $[a,b]$ included in the Wigner sea $[-2,2]$. 
We want to compute the distance $D$ between this subspace and the subspace spanned by 
the pertubed eigenvectors corresponding to all eigenvalues contained in $[a-\delta,b+\delta]$. Using the above formula, we find that in the $N \to \infty$ limit, $D$ tends to a finite limit as soon as $\delta > 0$:
\be\label{det(G^TG)}
D(a,b;\delta) \approx \frac{\epsilon^2}{2\int_a^b \rho(\lambda) d\lambda} \int_a^b d\lambda \int_{[-2;2]\setminus[a-\delta;b+\delta]} d\lambda' \frac{\rho(\lambda)\rho(\lambda')}{(\lambda - \lambda')^2}.
\ee
We checked this formula using numerical simulations, with very good agreement for different values of $a,b$ and $\delta$, and for $\epsilon$ up to 
$0.1$.\footnote{One can be much more precise and compute the full distribution of all 
singular values, giving an indication of their scatter around the mean position $\langle s \rangle$ \cite{usinprep}.} It is interesting to study the above expression in the double limit $\delta \to 0$ and $\Delta = b-a \to 0$. One finds:
\begin{equation}\label{cases}
D(a,a+\Delta;\delta) \approx \begin{cases} 
\epsilon^2 \frac{\rho(a) \ln(\Delta/\delta)}{\Delta} & \text{if $\delta \ll \Delta \ll 1$},\\
\epsilon^2 \frac{\rho(a)}{\delta}  &  \text{if $1 \gg \delta \gg \Delta$}. 
\end{cases} 
\end{equation}
This last expression shows that when the width $\Delta$ of interval $[a,b]$ tends to zero, the corresponding eigenvectors are scattered in a region of 
width $\delta$ much larger than $\Delta$ itself as soon as $\epsilon \gg \sqrt{\Delta}$. It also shows that for fixed $\Delta$, the distance $D$ diverges logarithmically
when $\delta \to 0$. This is a consequence of the pseudo-collisions that occur between eigenvalues close to the boundaries of the interval $[a,b]$. 
When $\delta > 0$, these pseudo-collisions are avoided and $D$ remains finite. When $\delta = 0$, a more precise analysis is needed. One can show that 
in the limit $N \to \infty$, the following result holds \cite{usinprep}:
\be
D(a,b;\delta=0) \approx \ln N \,\, {\epsilon^2} \,\, \frac{\rho(a)^2+\rho(b)^2}{2\int_a^b \rho(\lambda) d\lambda } + A(a,b)
\ee
where $A(a,b)$ is a constant that can be explicitely computed, and involves the well known two-point function $g(r)$ that describes the level-level correlations in the GOE. 
The $\ln N$ term can be guessed from the logarithmic behaviour of $D$ when $\delta \to 0$, since one indeed expects the divergence to be cut-off when $\delta$ becomes of the order of the 
level spacing, i.e. $\delta \sim (N \rho)^{-1}$. As a side remark, we note that Eq. (\ref{cases}) predicts that a fraction $\propto \delta^{-1}$ of the original eigenspace gets shoved away at distances larger than $\delta$ 
(in eigenvalue space). In the context of the non adiabatic evolution of a quantum system \cite{Wilkinson}, this implies that the energy of the system makes jump with a power-law distribution of sizes, such that all moments of order $q \geq 1$ 
diverge. This means that under an extreme non-adiabatic process, the energy is not diffusive but rather performs a ``Cauchy flight'' (i.e. a L\'evy flight with a tail exponent equal to $2$). When the
perturbation varies on a finite time $\tau$, one expects the tails of the jump process to be truncated beyond $\delta \sim \hbar/\tau$ \cite{Wilkinson}.

{\it{Eigenvector stability for covariance matrices.}} As mentionned in the introduction, covariance matrices are crucial to many problems in theoretical and applied finance. Empirically, the 
conundrum is the following: on the one hand, one needs sufficiently long time series in order to reduce the statistical noise in the determination of these matrices, but on the other hand, one {\it a priori} expects that covariance matrices do evolve with time. If the measurement time is too short, the empirical covariance matrix will appear to evolve with time, but this may just be due to the measurement noise that is
not reproducible from one period to the next. If the measurement time is too long, one may miss important correlation shifts and get exposed to unwanted sources of risk. 
The theory we develop here provides a precise estimate of the amount of eigenspace instability induced by measurement noise. Any extra
dynamics of the eigenvectors is therefore attributable to a genuine market evolution.

\begin{figure}[h!btp] 
		\center
		\includegraphics[scale=0.20]{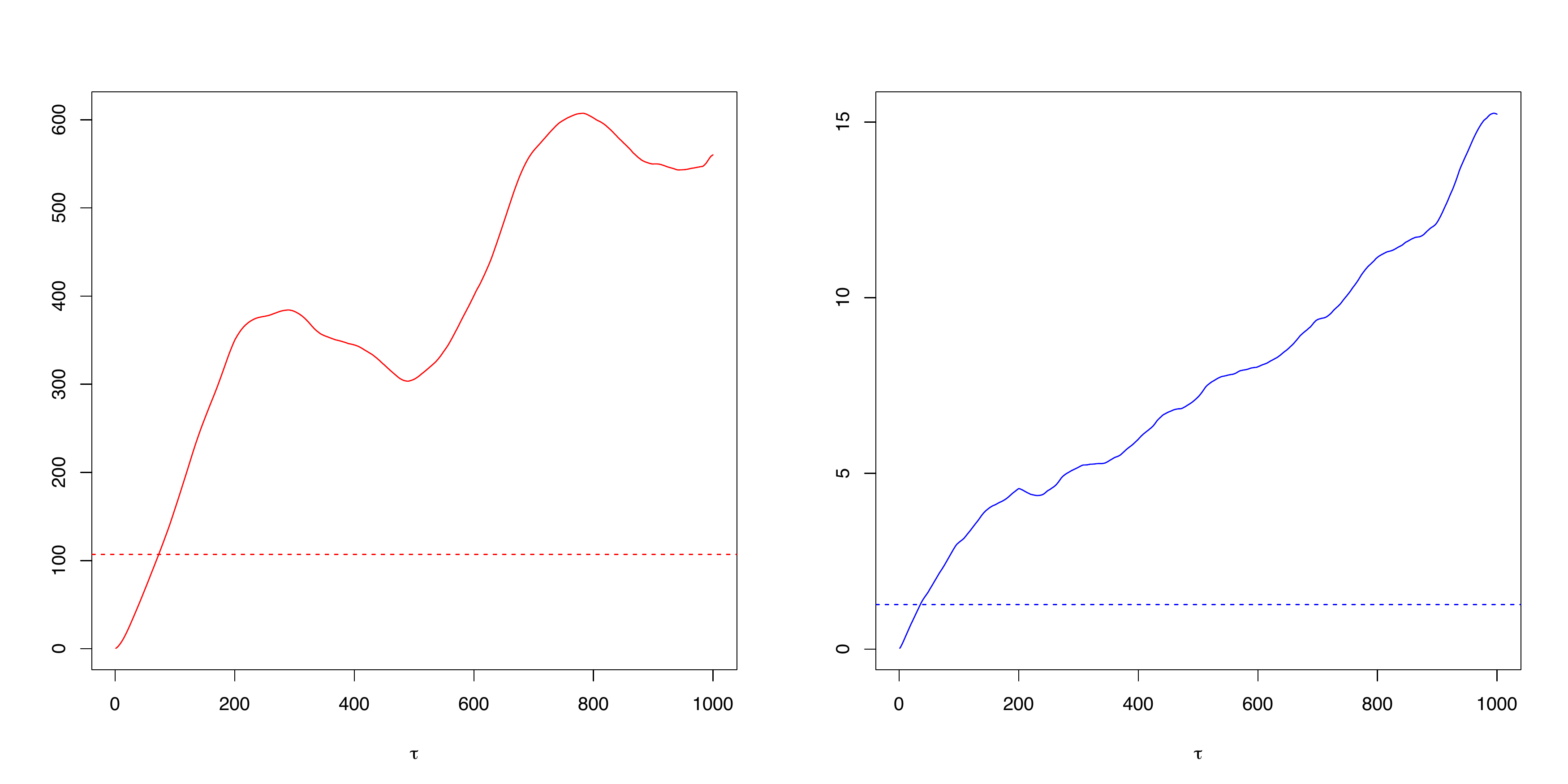}
                \caption{Empirical variograms $\langle(\lambda_i^{s} - \lambda_i^{t})^2\rangle_{|t-s|=\tau}$ for $i=1$ (left) and $i=2$ (right). The 
                horizontal dotted lines correspond to the theoretical prediction for a time independent correlation matrix.}         \label{fig1}
\end{figure}

Let $\left\{r_i(t)\right\}_{1\leq i \leq N,1\leq t \leq T}$ be the set of returns (for example on a daily time scale) of $N$ assets over a period of length $T$, which can be seen as a $N\times T$ 
random rectagular matrix, $R_N$. Let us assume that these returns are stationary gaussian variables with zero mean and a (true) covariance matrix  ${\bf C}$. 
The empirical (or ``sample'') covariance matrice is defined as: ${\bf E} := \frac1T R_N^\dagger  R_N$. When $T \to \infty$, ${\bf E} \longrightarrow {\bf C}$ since we assume for 
now that the underlying return process is stationary. For finite $T$ however, one has \cite{Krakow}:
\be \label{pertubSigma}
{\bf E} = {\bf C} + {\bf \mathcal{E}}, \quad {\rm {with}} \quad \mathcal{E}_{ij} = \frac{1}{T} \sum_{t=1}^{T} r_i(t) r_j(t) - C_{ij}.
\ee
Thus the sample covariance matrix ${\bf E}$ is a perturbed version of ${\bf C}$. Each of the $\mathcal{E}_{ij}$ has $0$ mean and the covariance structure of the entries of $\mathcal{E}$ is given by 
$\left\langle\mathcal{E}_{ij} \mathcal{E}_{k\ell}\right\rangle =  (C_{ik} C_{j\ell} + C_{i\ell} C_{jk})/T$. Using the same framework as above, one can calculate the distance (or overlap) between 
the top $P$ eigenvectors of the true correlation matrix ${\bf C}$ and the top $Q$ eigenvector of the empirical 
correlation matrix ${\bf E}$. [We focus on the top eigenvalues and eigenvectors because these represent the most risky directions in a financial context.] 
Provided $T$ is large enough for the above perturbation theory to be valid, and upon averaging over the measurement noise:
\be \label{noiseVecteur}
D(P,Q) = \frac{1}{2TP} \sum_{i=1}^{P} \sum_{j=Q+1}^{N} \frac{\lambda_i \lambda_j}{(\lambda_i - \lambda_j)^2}, 
\ee  
where the $\lambda_i$s are the eigenvalues of ${\bf C}$, in decreasing order. One can similarly define the distance between the eigenspaces of two independant sample covariance matrices 
${\bf E}^s$ and ${\bf E}^t$ (determined on two non overlapping time periods), with:
\be
E_{ij}^u = \frac1T \sum_{v=1}^T r_i(v+u) r_j(v+u), \quad  u=(t,s), \quad |t-s|>T
\ee
In this case, the above formula Eq. (\ref{noiseVecteur}) is simply multiplied by a factor $2$. In practice of course one does not know the true matrix ${\bf C}$, 
and whether it evolves in time or not. If ${\bf C}$ was time {\it independent}, the measurement noise on its eigenvalues should be such that, for all $i$:
\be
\left\langle\left(\lambda_i^{s} - \lambda_i^{t}\right)^2\right\rangle_{|t-s|>T} \approx \frac{4 \lambda_i^2}{T}. 
\ee 
where the $\lambda_i$ are the eigenvalues of the matrix ${\bf C}$ measured empirically using the whole period of time and where $\langle \cdot \rangle_{|t-s|>T}$ denotes an empirical average over all 
$s,t$ such that ${|t-s|>T}$. We determined the empirical variograms $\langle(\lambda_i^{s} - \lambda_i^{t})^2\rangle_{|t-s|=\tau}$ for $i=1,2$, using daily returns of $N=204$ stocks 
of the Nikkei index in the period $2000-2010$. The result is shown in Fig. \ref{fig1}, which clearly shows that there is a genuine evolution of the 
eigenvalues of ${\bf C}$ with time. For the top eigenvalue, this is a well known effect \cite{Krakow}: both the volatility of individual stocks and the average correlation between stocks are time dependent. We 
see that the same is true for smaller eigenvalues too, reflecting the instability of intra-sector correlations. 

What about the eigenvectors ? One could be in a situation where the eigenvectors keep a fixed direction through time while eigenvalues are moving around. 
But if the eigenvalues themselves are evolving with time, the above formulas need to be upgraded. Let us assume that the true covariance matrix ${\bf{C}}_t$ has time dependent eigenvalues 
$\lambda_1^t,\dots,\lambda_N^t$ but with constant eigenvectors that will be denoted $|\phi_1\rangle, \dots, |\phi_N\rangle$ as above. For times $s<t$ with $|t-s|\geq T$, we define the overlap matrix 
${\bf{G}}^{s,t}$ as: $G_{ij}^{s,t} = \langle \phi_i^{s} | \phi_j^{t} \rangle$. Under the assumption that the eigenvalues are varying sufficiently slowly  with time, one now finds that:
\begin{align}
D&(P,Q;s,t) = - \frac{1}{2P}  \left\langle \ln |\det({{\bf G}^{s,t}}^\dagger{\bf G}^{s,t})| \right\rangle\nonumber \\ 
&\approx  \frac{1}{2TP} \sum_{i=1}^{P} \sum_{j=Q+1}^{N} \left(\frac{\lambda_i^s \lambda_j^s}{(\lambda_i^s - \lambda_j^s)^2}  
+ \frac{\lambda_i^t \lambda_j^t}{(\lambda_i^t - \lambda_j^t)^2} \right). \label{timeDep}
\end{align}
Up to corrections of order $T^{-3/2}$, one can replace in the above formulas the $\lambda^{s,t}$ by their empirical estimates. 
We finally compute the theoretical distance $D_{th}(P,Q,\tau)$ as an average over all $s,t$ such that $|t-s|=\tau$ of the above quantity. 

\begin{figure}[h!btp] 
		\center
		\includegraphics[scale=0.28]{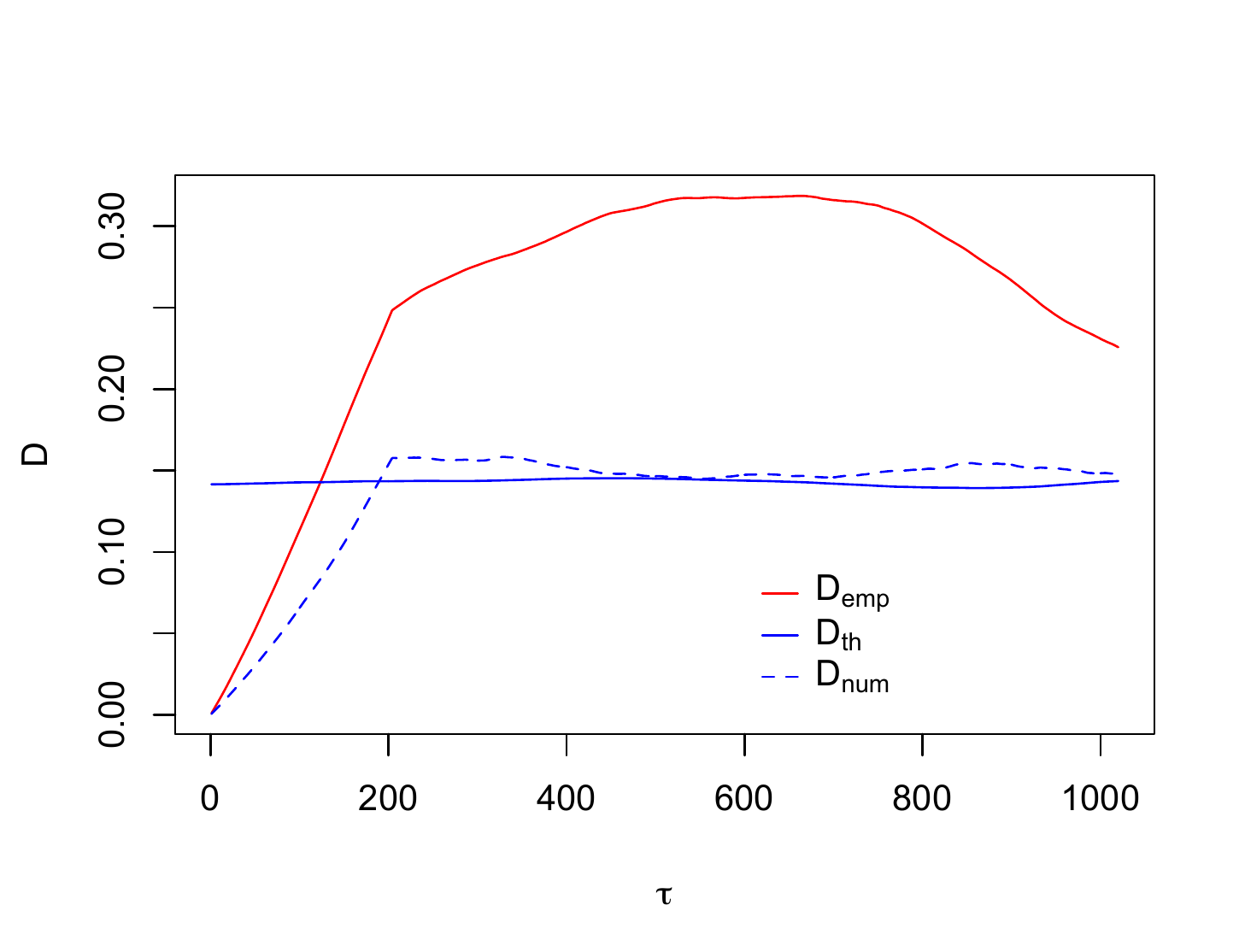}
                  \includegraphics[scale=0.28]{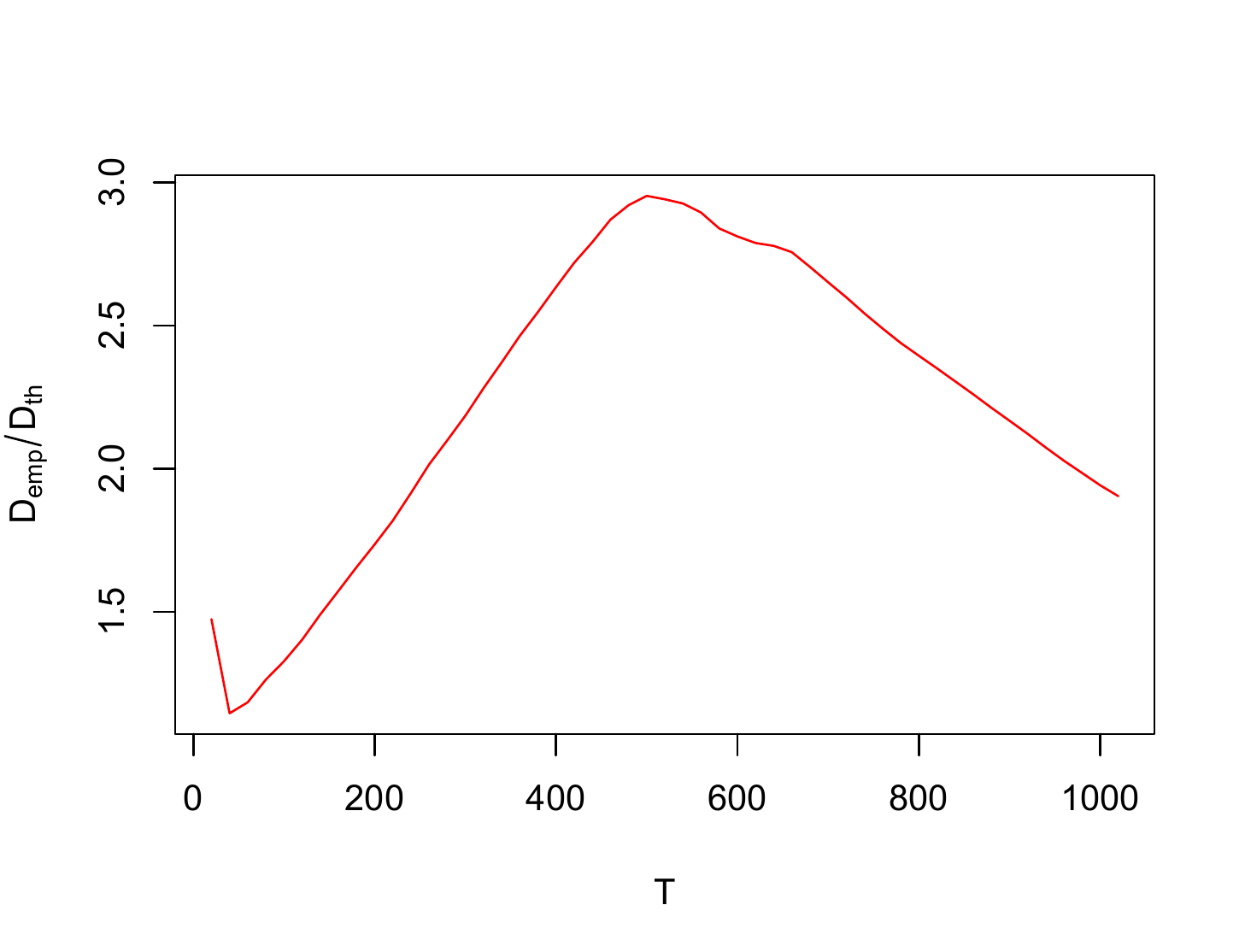}
        \caption{Left: $D_{th}, D_{num}, D_{emp}$ as a function of $\tau$ with $P=5, Q=10$ and $T=204$ for $N=204$ stocks of the Nikkei index in the period 2000-2010. Note that $D_{emp}$ reaches a maximum for $\tau \approx 500$, which is significantly smaller than the random benchmark $D_{RMT} \approx 0.83$ for $P=5, Q=10, N=204$. Right: The ratio $D_{emp}/D_{th}$ as a function of $T$, for $\tau=T$, $P=5, Q=10$, revealing the same maximum for $T^* \approx 500$.}   \label{fig2}      
\end{figure}

We now compare our null hypothesis formula, Eq. (\ref{timeDep}) with (a) an empirical determination of $D_{emp}(P,Q,\tau)$ using financial data and (b) a numerical determination of $D_{num}(P,Q,\tau)$ using synthetic 
time series of returns that abide to the hypothesis of a covariance matrix ${\bf C}_t$ with {\it fixed} eigenvectors, but time dependent eigenvalues. We choose an arbitrary (but fixed) set 
of orthonormal vectors $|\psi_1\rangle, \dots, |\psi_N\rangle$ and define ${\bf C}_t$ as ${\bf C}_t = \sum_{i=1}^{N} \lambda_i^{t} |\psi_i\rangle \langle \psi_i|$,
where the $\lambda^{t}$ are the empirical eigenvalues obtained on the financial return time series. We then use ${\bf C}_t$ to generate synthetic gaussian 
multivariate returns $\left\{r_i(u)\right\}$. We show the corresponding results in Fig. \ref{fig2}, with the choice $P=5,Q=10$, as a function of $\tau$ and for $T=204$ days. We
conclude that (i) the theoretical formula Eq. (\ref{timeDep}) is indeed in very good agreement with the numerical results obtained with synthetic data: $D_{num} \approx D_{th}$; whereas (ii) the financial data clearly 
departs from the null hypothesis of constant eigenvectors, since $D_{emp} > D_{th}$. The same conclusion holds for different values of $P,Q$. We also show the ratio $D_{emp}/D_{th}$ for $\tau=T$, as a function of $T$. 
This plot reveals a marked maximum around $T^* \approx 2$ years, suggesting that the correlation matrix has some true dynamical evolution with a mean reversion time around $T^*$, with important consequences for risk analysis.

As a conclusion, we have developed general tools to describe the dynamics of eigenvectors under the influence of random perturbations. We
obtained exact results in the context of random matrices in the GOE ensemble and correlation matrices, with in mind applications to financial risk, but hopefully the ideas and methods introduced here can be used in a much broader context.


\begin{thebibliography}{20}

\bibitem{Verdu} A. Tulino, S. Verd\`u, {\it Random Matrix Theory and Wireless Communications}, Foundations 
and Trends in Communication and Information Theory, {\bf 1}, 1-182 (2004).
\bibitem{Handbook} G. Akemann, J. Baik, Ph. Di Francesco, {\it The Oxford Handbook of Random Matrix Theory}, Oxford University Press (2011).
\bibitem{Simon} B. D. Simon, P. A. Lee and B. A. Altshuler, Phys. Rev. Lett. 70 (1993). 
\bibitem{Wilkinson} M. Wilkinson, J. Phys. A: Math. Gen. 21 (1988) 4021-4037. Phys. Rev. A 41, 4645-4652 (1990). 
\bibitem{Review} J.-P. Bouchaud, M. Potters, {\it Financial Applications of Random Matrix Theory: a short review}, in \cite{Handbook}. 
\bibitem{RSVD} J.-P. Bouchaud, L. Laloux, M. A. Miceli and M. Potters, Eur. Phys. J. B {\bf 55} (2007) 201.
\bibitem{usinprep} R. Allez, J.-P. Bouchaud, in preparation.
\bibitem{Krakow} M. Potters, J.-P. Bouchaud, L. Laloux, Acta Phys. Pol. B {\bf 36}, 2767 (2005).

\end{thebibliography}
\end{document}